\documentstyle[epsfig,psfig]{article}
\begin{document}
\title{Shot noise and higher current moments in dimensions 2, 3
\& 4}

\author{I. Trav\v enec \\ 
Institute of Physics, Slovak Academy of Sciences,
D\'ubravsk\'a cesta 9, \\
84511 Bratislava, Slovakia}

\maketitle 

\begin{abstract}
Shot noise and higher current moments $T_M=\langle {\rm Tr} (t^{\dagger}t)^M
\rangle$ are studied within the Anderson model of disordered conductors in 
dimensions $d= 2, 3$ and 4. Here $t$ denotes a transmission matrix. We 
calculate the conductance $G=T_1$, shot noise $S=T_1-T_2$, Fano factor 
$F=S/G$ and ratios $C_M=T_M/G$, and we show indications that all of them 
are as good order parameters as the conductance itself. 
In the limit of infinite system size, two limiting values of $F$ and $C_M$
are found; the stable one in the metallic regime, and the unstable one,
characterizing the critical point for $d > 2$. 
We present analytical expressions for both limiting values, together with 
a compact formula for current cumulants at 3D criticality. 
Our data confirm also Nazarov's microscopic theory [PRB {\bf 52}, 4720 
(1995)], as we show numerically for special linear combinations of $T_M$.
\end{abstract}

\renewcommand{\leftmark}
{I. Trav\v enec: Shot noise and higher current moments}

\section{Introduction}

Much effort has been devoted to the Anderson localization \cite{and}.
The transport in disordered systems exhibits many universal features, like
universal conductance fluctuations (UCF) in the metallic regime, or 
$\ln G$ - type distribution in the insulator.

Theoretical approach concentrated mainly, but not exclusively, on 
one-dimen\-sional (1D) models. 
The type of scaling of the conductance distribution was investigated 
in 1D \cite{dey} and numerically in 3D \cite{sle}.
Muttalib et al. \cite{mut} analyzed and generalized the DMPK equation also
beyond quasi-1D and calculated the distribution function of $G$ in all 
regimes. They tried to apply the generalized DMPK also to localized 
states \cite{mamu}; in the metallic case it was applied successfully before
\cite{rmt}. Beenakker and B\"uttiker \cite{bebu} also studied 
the metallic transport in dimension $d>1$ and we will utilize their results 
for all current moments. 

Besides the UCF, i. e. the second cumulant of $G$ (or $var\ G$), 
numerous other exact quantitities where 
proposed in the metallic regime (see this Intro and Section 2), but only few 
in the other regimes. The main purpose of this work is to calculate a 
complete set of quantities at 3D metal-insulator transition (MIT) exactly. 
It contributes to theoretical description of the criticality from different 
side than microscopic theory, RMT, non-linear $\sigma$-model, DMPK, etc.

Besides the Landauer's conductivity 
$G={\rm Tr}\ {\cal T}$ (in units $2e^2/h$), and shot noise power 
$S = {\rm Tr}[{\cal T}-{\cal T}^2]$, 
where ${\cal T}=t^{\dagger}t$, also the universal properties of higher 
order current moments in terms of ${\cal T}^M$ were observed \cite{high}. 
The third current cumulant was given analytically also in the limit 
of frequency and temperature both zero:
\begin{equation}
  \label{eq:1}
  \langle\langle I^3\rangle \rangle \ = \langle I^3\rangle - 3\langle I^2
  \rangle \langle I\rangle  + 2\langle I\rangle^3  = \langle I\rangle \ 
\frac {\langle{\rm Tr} [{\cal T}(1- {\cal T})(1-2{\cal T})]\rangle}
{\langle{\rm Tr}\ {\cal T}\rangle}
 = \frac {\langle I\rangle}{15}
\end{equation}
$\langle ...\rangle$ means ensemble averaging.
This formula holds for large systems in diffusive (metallic) regime and 
we will reproduce it later. 

Let us denote the higher current moments $T_M=\langle{\rm Tr}\ 
[{\cal T}^M]\rangle$ . These
quantities were analyzed in the works \cite{pendry} and \cite{kirk}.
The exact solution for normalized moments on 1D Anderson chain 
\begin{equation}
  \label{eq:2}
 C_M=\frac{T_M}{T_1}=\frac{ \Gamma(M-\frac{1}{2})^2}{\Gamma(M)^2 
\Gamma(\frac{1}{2})^2}
\end{equation}
was confirmed by numerical simulations.
This formula applies exclusively to pure 1D. Analogical exact 
expressions for higher dimensions were given in \cite{bebu} for the metallic 
transport, Eqs. (\ref{eq:A2}) or (\ref{eq:8}), and we will add those 
for 3D critical regime. 

Let us make a note on relationship of cumulants and moments. Compact 
expressions for charge (i. e. also current) cumulants $\ll Q^k\gg$ in 
diffusive regime were derived in \cite{ll}. These cumulants can be 
calculated with help of a characteristic function
\begin{equation}
  \label{eq:3}
\ln \chi(u)=\sum_{k=1}^\infty \frac{u^k}{k!}\ll Q^k\gg
\end{equation}
where $\chi$ is Fourier transformed probability distribution of charges.
Assuming mutual independence of channels they found
\begin{equation}
  \label{eq:4}
\ll Q^k\gg \ \propto \sum_j \Bigl({\cal T}(1-{\cal T})\frac{d}{d{\cal T}}
\Bigr)^{k-1}{\cal T}\big|_{{\cal T}={\cal T}_j}
\end{equation}
where the sum over eigenvalues means trace. We get shot noise as 
the second cumulant, proper part of Eq. (\ref{eq:1}) as the third 
one, etc. Thus the current cumulants and moments are connected  
in a different way than e. g. those ones of conductivity.

The moments $T_M$ themselves become usually either zero (insulator) 
or infinite (metal) in the large system limit, except criticality. 
But we will show, that both their ratios and appropriately chosen linear 
combinations will have non-trivial finite values. 
The limiting $C_M$ ratios have the pleasant property that they are 
independent of boundary conditions.

The shot noise $S=T_1-T_2$ and Fano factor $F=1-C_2$ of Anderson model have 
been studied in \cite{shot12} for dimensions $d=Q1D,2$ and in \cite{shot3}
for $d=3$. Q1D means quasi-1D systems, i. e. stripes of fixed, but large 
enough width and even much larger length. We continue in 2D and 3D with 
larger systems and statistics, and we add both $d=4$ 
and higher current moments $M \ge 3$. Under simple 
assumptions on the distribution function of Lyapunov exponents, one gets 
the limiting values for $C_M$ both in diffusive and critical (3D) regimes. 
We will also confirm the validity of Nazarov's theory \cite{naz} in diffusive 
case for higher moments and in $d=4$; for $M=2$ and $d=Q1D,2,3$ it was 
already shown in \cite{shot12} and \cite{shot3}, though less precisely 
because of smaller system sizes and thus larger disturbing ballistic effects. 

There is a great amount of literature on shot noise, e. g. \cite{bebu},
\cite{high}. The Poissonian value
$F \to 1$ corresponds to the strongly localized (insulating) regime,
the well known 1/3 suppression $F=1/3$ is typical for the metallic regime.
If $d > 2$ we get yet another value $F_C$, which marks the MIT. All higher 
moments will show qualitatively the same behaviour.

In recent paper \cite{gala} even the fourth cumulant, or the combination
$C_3 - C_4$ was introduced. Such quantities can be calculated in general. 

The aim of this work is to show that $C_M$ and special linear combinations 
of $T_M$ converge to universal values in the large system size limits; 
to observe two non-trivial limiting $C_M$ values for $d > 2$; to give 
them analytically (except 4D criticality) and to prove them numerically.

The paper is organized as follows. In this Intro and in Section 2.1 we 
recall some results from papers,
treating higher current moments or cummulants, especially in metallic
regime, and the first steps towards 3D criticality. Then we add theoretical
predictions for any normalized moment and cummulant at 3D MIT. In the next
Section 2.2 we briefly introduce Nazarov's microscopic model, yielding even
more information about the moments in metallic regime. In Section 3 
we confirm virtually any given analytical formula for current moments
and cummulants by numerical simulations (except 1D and Q1D).

\section{Model, Numerical Method and Theory}

Let us consider the Anderson model with the Hamiltonian
\begin{equation}
  \label{eq:5}
 H=\sum_n \epsilon_n |n><n| + \sum_{[nn']} |n><n'|+|n'><n|
\end{equation}
where the diagonal disorder $\epsilon_n$ is taken randomly from $<-W/2,W/2>$, 
i.e. it has uniform (box) distribution, and $[n,n']$ are nearest neighbor 
sites of a $d$-dimensional cubic lattice. Concerning numerical calculations, 
the transfer matrix method of the calculation of moments and especially 
the way how to deal with larger systems was described in \cite{pendry}. 

\subsection{Limiting values of Fano factor, $C_M$ and current cumulants}

Lyapunov exponents $\lambda_n$ are related to the eigenvalues ${\cal T}_n$ 
of ${\cal T}$ as follows:
\begin{equation}
 \label{eq:6}
 {\cal T}_n=\cosh^{-2}(\lambda_n)
\end{equation}
It has long been known, that the smallest (most important) $\lambda_n$ 
exponents are distributed equidistantly in large metallic systems, i. e. 
that their distribution function $P(\lambda)$ is constant. In the work 
\cite{bebu} this was used to calculate general diffusive $C_M$. Constant
density was observed numerically and approved by various analytical
treatments. Numerical data, however, indicate that in the critical regime 
(3D, 4D) $P(\lambda)$ reaches the origine, and it is not constant.
To generalize the results, we can propose that at least for small
$\lambda$, the $P(\lambda) \propto \lambda^\alpha$. We know 
already that $\alpha=0$ in metallic regime, irrespective of dimension. 
Replacing the trace, i. e. summation of eigenvalues, by an integral over 
$P(\lambda) d\lambda$ - which works for large enough systems - we get:
\begin{equation}
 \label{eq:7}
C_M(\alpha)={\int_0^\infty \lambda^\alpha \cosh^{-2M}(\lambda) d\lambda \over 
 \int_0^\infty \lambda^\alpha \cosh^{-2}(\lambda) d\lambda}
\end{equation}
Now let us summarize the metallic regime. The one-third-suppression 
$F=1-C_2=1/3$ was confirmed in \cite{bebu}. We can also reproduce the 
Eq. (\ref{eq:1}), namely $1-3C_2+2C_3=1/15$. The term proposed 
in \cite{gala} is nothing but the fourth cumulant: $F-6(C_2-C_3)+6(C_3-C_4)
=-1/105$. And generally for the Fano-factor-like quantities we get 
$C_M-C_{M+1}=C_M/(2M+1)$, with 
\begin{equation}
 \label{eq:8}
C_M(0)=\frac{4^{M-1}(M-1)!^2}{(2M-1)!}
\end{equation}
which is yet another practical form of Eq. (\ref{eq:A2}), \cite{bebu}.
An analogy of Eq. (\ref{eq:7}) was proposed already in \cite{pendry}, just
with unspecified distribution function.

In \cite{pet} it was shown that $\alpha=1$ in 3D at MIT. This was utilized
in \cite{shot3} to calculate the limiting value of Fano factor
in the critical regime. We can give this value exactly:
$F_C=1-C_2(1)=1/3+1/(6\ln 2)$. In Appendix A the integrals of interest 
are solved analytically. Useful  values of $C_M(\alpha)$ are summarized 
in Table \ref{tab:1}. 

The $\alpha=1$ is correct not only for $P(\lambda)$, 
but also for the distribution of the first (smallest) Lyapunov exponent 
$P(\lambda_1)$, even with the same slope. That is why the fits are perfect
for all moments - for numerically treated finite systems the higher moments
depend almost exclusivelly on $P(\lambda_1)$.
The possibility of $\alpha=2$ was proposed in \cite{mahe} for 4D at MIT;
we will come to this point later.

Let us return to the current cumulants. The generating function $\ln\chi_
{\alpha}(u)$ from
Eq. (\ref{eq:3}) was calculated for the $\alpha=0$ case in \cite{ll}:
\begin{equation}
 \label{eq:9}
\ln\chi_0(u)=Q_0\int_0^{\infty}d\lambda\ln\Bigl(\frac{e^u-1}{\cosh^2(\lambda)}
+1 \Bigr)=Q_0\ {\rm arcsinh}^2\sqrt{e^u-1}
\end{equation}
where the charge $Q_0$ involves constants and physical units, see \cite{ll},
and it is non-universal, depending on model parameters like disorder $W$. 
Using Eq. (\ref{eq:3}) they calculated the cumulants $\ll Q\gg/Q_0=1$, 
$\ll Q^2\gg/Q_0=1/3$, $\ll Q^3\gg/Q_0=1/15$, see Eq.(\ref{eq:1}), 
$\ll Q^4\gg/Q_0=\ll Q^5\gg/Q_0=-1/105$, see above, etc.
In the case $\alpha=1$ we get a rather formal formula (Appendix A)
\begin{eqnarray}
 \label{eq:10}
 \nonumber
\ln\chi_1(u) & = & Q_0'\int_0^{\infty} \lambda \ln\Bigl(\frac{e^u-1}
{\cosh^2(\lambda)}+1 \Bigr) d\lambda
=-2Q_0'\int_0^{t^*}\Bigl(t\ln 2-L(t)\Bigr)dt= \\ 
& = & Q_0'\ln 2 \ {\rm arcsinh}^2\sqrt{e^u-1}+2Q_0'\int_0^{t^*}L(t)dt
\end{eqnarray}
where $t^*=i\ {\rm arcsinh}\sqrt{e^u-1}$ and the Lobachevsky function,
see \cite{grry}:
\begin{equation}
\label{eq:11}
L(z)=-\int_0^z \ln(\cos(x)) dx = \frac{z^3}{6}+\frac{z^5}{60}+\frac{z^7}
{315}+{\cal O}(z^9)
\end{equation}
We can calculate the cumulants at 3D criticality, taking again the
derivatives of $\ln\chi_1(u)$ at $u \to 0$: 
\begin{eqnarray}
\label{eq:12}
\ll Q\gg & = & Q_0'\ \ln 2 = T_1 \\
\ll Q^2\gg & = & Q_0'\ \Bigl( \frac{1}{3}\ln 2 +\frac{1}{6} \Bigr) 
= T_1-T_2 \\
\ll Q^3\gg & = & Q_0'\ \Bigl( \frac{1}{15}\ln 2 +\frac{2}{15} \Bigr) 
= T_1-3T_2
+2T_3 \\
\ll Q^4\gg & = & Q_0'\ \Bigl(-\frac{1}{105}\ln 2 +\frac{11}{210} \Bigr)
= T_1-7T_2+12T_3-6T_4 
\end{eqnarray}
We added relations in terms of $T_M$ from Eq. (\ref{eq:4}).
The above-mentioned value $F_C=\ll Q^2\gg/\ll Q\gg$.

\begin{table}[b]
\caption{Selected values of $C_M$ from Eq. (\ref{eq:7}). Analytical values
of $C_M(1)$ can be deduced from Appendix A}
\label{tab:1}\renewcommand{\arraystretch}{1.5}
\begin{tabular}{llllllll} \hline
$\alpha$ & $C_2$ & $C_3$ & $C_4$ & $C_5$ & $C_6$ & $C_7$ & $C_8$ \\ \hline
0 & ${2\over 3}$ & ${8\over 15}$ & ${16\over 35}$ & ${128\over 315}$
& ${256\over 693}$ & ${1024\over 3003}$ & ${2048\over 6435}$\\
1 & 0.426217 & 0.268839 & 0.196084 & 0.154259 & 0.127120 & 0.108094
& 0.0940195 \\ \hline
\end{tabular}
\end{table}

\subsection{Diffusive $T_M$ values from Microscopic theory}

The diffusive contribution to eigenvalues of ${\cal T}$ was calculated 
in \cite{naz}.
Nazarov introduced a function $\delta F(\phi)$, and he showed that
$P(\lambda)$ is proportional to real part of $\sin\phi\ \delta F(\phi)$
at special complex $\phi$ and that

\begin{equation}
 \label{eq:16}
\delta F(\phi)= {\rm Tr}\Bigl({{\cal T}\over {\it 1}-{\cal T}\sin^2\
{\phi\over 2}}\Bigr)=-{2\phi\over \sin\phi} \sum_s{1\over s^2-\phi^2}
\end{equation}

Here the Cooperon mode energies $s^2=\pi^2(n_1^2+n_2^2+...+n_d^2)$, 
$n_1=1,2,...$, $n_2=0,1,...$
, ..., $n_d=0,1,...$ for simple $d$-dimensional 
cube with hard wall boundary 
conditions. The index $n_1$ refers to 
one chosen transfer direction. The minor changes for samples with
periodic boundary conditions were described e. g. in \cite{bra}. There is 
analyzed the 3D case of $G=T_1$, because the sum on the rhs of Eq. 
(\ref{eq:16}) diverges for $d \ge 2$ and a cut-off on wave vectors' length 
is useful.
Some other details can be found in \cite{it}. $\phi$ is the global phase shift
between left and right side of the sample and it also plays the role 
of potential difference \cite{naz}. We will use it to get 
the moments 
of ${\cal T}$. Taking derivatives with respect to $\phi^2$ at
 zero, details 
in Appendix B, we get the following linear combinations
 $R_M$

\begin{eqnarray}
\label{eq:17}
R_1=-T_1 & = & 2 S_1 \nonumber \\
R_2=-T_2+{2\over 3}T_1 = S-{G\over 3} & = &  8 S_2 \nonumber \\
R_3=-T_3+T_2-{2\over 15}T_1 & = & 32 S_3   \\ 
R_4=-45T_4+60T_3-18T_2+{4\over 7}T_1 & = & 5760 S_4 \nonumber \\ 
R_5=-315T_5+525T_4-245T_3+{85\over 3}T_2-{2\over 9}T_1 & = & 161280 S_5 
\nonumber \\
R_6=-14175T_6+28350T_5-17955T_4+3840T_3-186T_2+{4\over 11}T_1 & = 
& 29030400 S_6 \nonumber
\end{eqnarray}

We introduced the sums $S_M=\sum_s s^{-2M}$. The way how to calculate
these sums efficiently in any dimension is also in Appendix B. The first
two of these equations were derived and the second one compared to numerical 
simulations
 in \cite{shot12} and \cite{shot3}. The values on rhs are summarized
in Table \ref{tab:2}. 

Recall that for $d \ge 2$ the conductivity $T_1 \to \infty$ and for $d<2M$
the sums $S_M$ are finite, see Appendix B. One can divide all the equations 
by $T_1$ and get a homogeneous system for any $C_M$. Of course the results 
are the same as those from Eq. (\ref{eq:A2}). This means, that Nazarov's 
formula (\ref{eq:16}) incorporates the diffusive case $\alpha=0$ of Eq. 
(\ref{eq:7}), though the validity of the last goes beyond the mentioned 
restrictions on dimension (or $M$). Therefore it is usefull to calculate 
$1-C_2$ by numerical simulations at 4D and show that the 1/3 suppression 
holds, see Fig. \ref{fig:3} left. If we would like to derive this analytically
also for $d \ge 2M$, the cut-off $K$ could be introduced together with finite 
sums $S_M(K)$, where the summation is performed only over finite set of 
$\{n_1, ..., n_d\}$ such that the wave vector length $s < K$. Then we expect
$\lim_{K \to \infty} S_M(K)/S_1(K) = 0$ for all $M \ge 2$ and $d \ge 2$.

To proceed with comparing Eq. (\ref{eq:16}) and the attitude from previous
subsection, we would need the $\phi$ - dependence of the Lyapunov exponents
distribution $P(\lambda, \phi)$. We can conjecture that it is 
independent of $\lambda$, i. e. $P(\lambda,\phi)=G(\phi)$, and $G(0)=T_1$. 
Inserting this into the first expression for the generating function in
Eq. (\ref{eq:16}) and replacing the trace as in Eq. (\ref{eq:7}), we get:

\begin{equation}
 \label{eq:18}
\delta F(\phi) = G(\phi) \int_0^\infty \frac {\cosh(\lambda)^{-2} d\lambda}
{1-\cosh(\lambda)^{-2}\sin^2{\phi\over 2}} =  \int_0^\infty \frac {G(\phi) 
\ d \lambda}{\cosh(\lambda)+\cos{\phi}} = \frac{\phi}{\sin\phi}\ G(\phi)
\end{equation}
which is consistent with Eq. (\ref{eq:17}) at $\phi\to 0$, and comparison
to the rhs. of Eq. (\ref{eq:16}) yields $G(\phi)$. Nazarov \cite{naz} 
introduced yet another quantity $I(\phi)=\phi\ G$ and he gave a formula for
the Fano factor in terms of derivatives of $I$. We are now able to evaluate
them:

\begin{equation}
 \label{eq:19}
F = \frac{1}{3}\Bigl(1-\frac{2\ I'''(0)}{I'(0)}\Bigr)=\frac{1}{3}\Bigl(1-
\frac{12\ S_2}{S_1}\Bigr) 
\end{equation}
We can see again, that the one-third suppression is given by the infinite 
denominator $S_1$ rather than by zero numerator (constant, $\phi$
independent $G$), as previously assumed.

It also is interesting, that the $S_2$ value appears in the formula 
for UCF, e. g. \cite{it}, \cite{lsf}:
\begin{equation}
 \label{eq:20}
\langle G^2 \rangle - \langle G \rangle^2 = 12 S_2 = \frac {3} {2} 
\Bigl(S - \frac{\langle G \rangle}{3}\Bigr) = \langle G \rangle
- \frac{3}{2}T_2
\end{equation}

\begin{table}[b]
\caption{The sums $S_M$ with appropriate factors in dimensions
2, 3 and 4}
\label{tab:2}\renewcommand{\arraystretch}{1.5}
\begin{tabular}{llllll} \hline
d & 8 $S_2$ & 32 $S_3$ & 5760 $S_4$ & 161280 $S_5$ & 29030400 $S_6$ 
\\ \hline
2 & 0.123742 & 0.0387682 & 0.649759 & 1.77893 & 31.9117 \\
3 & 0.209369 & 0.0458759 & 0.699664 & 1.84186 & 32.4521 \\
4 & $\infty$ & 0.0573425 & 0.763394 & 1.91503 & 33.0469 \\ \hline
\end{tabular}
\end{table}

\section{Numerical simulations}

We calculated the mean values of $T_M$ and $C_M$, $M=1,2, ..., 8$ in
dimensions $d=2,3$ and 4 for various disorders and system lengths $L$. 
Typical ensembles were of $10^5$ elements for smaller systems and
$3\times 10^4$ for the greatest ones.

\subsection{Fano factor and $C_M$}

\begin{figure}[t]
\noindent
\includegraphics[width=.4\textwidth]{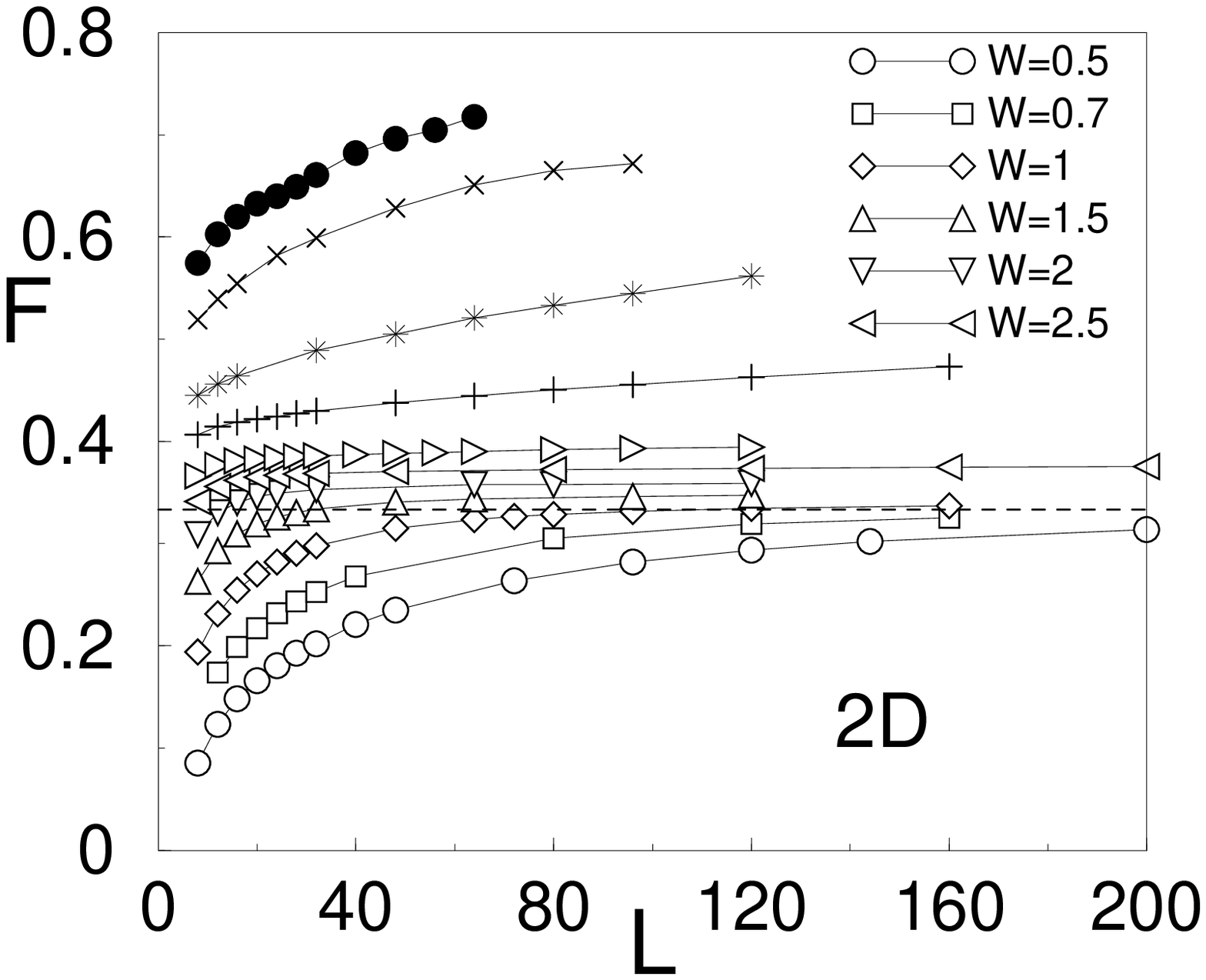}
\includegraphics[width=.4\textwidth]{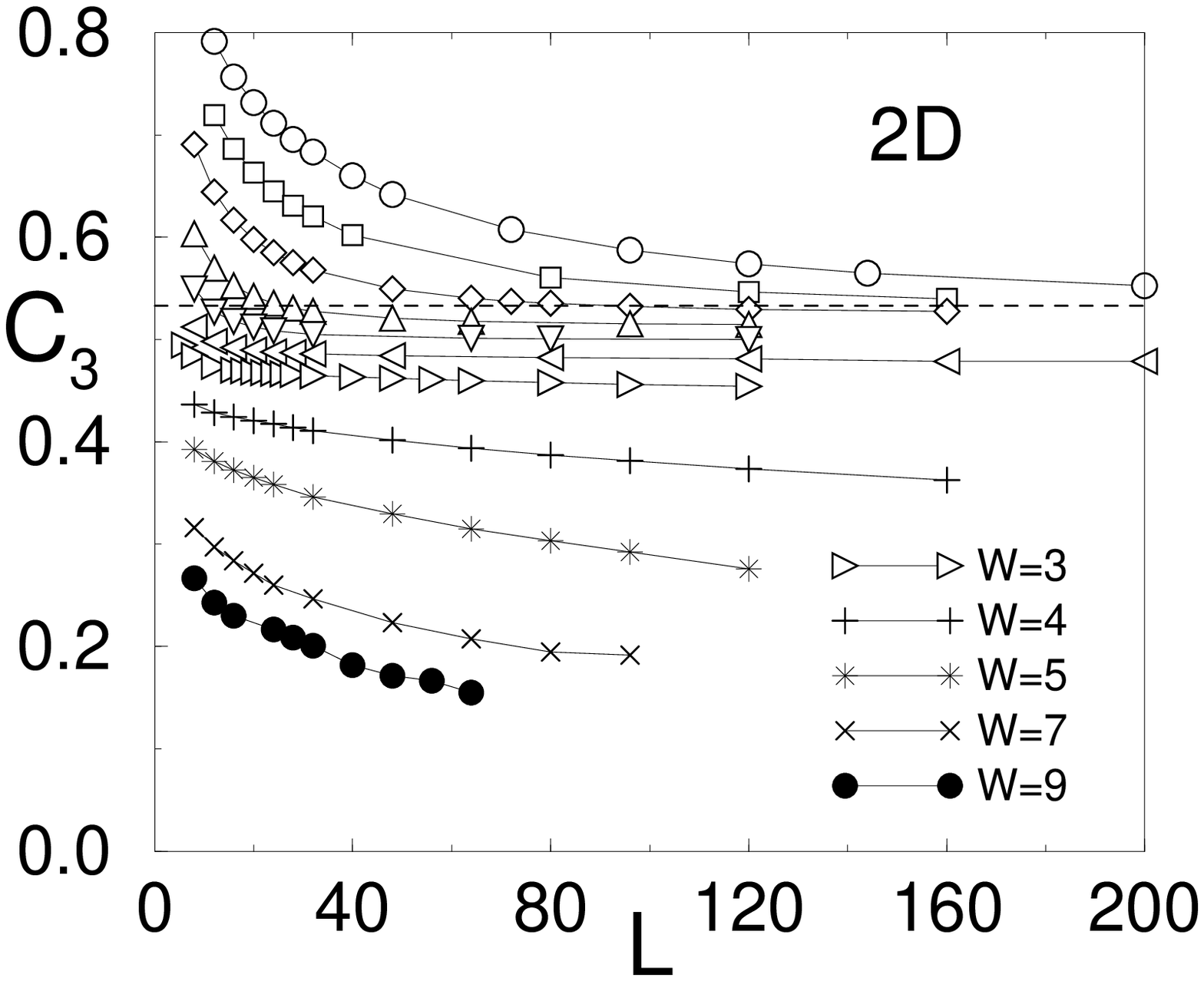}
\caption{Fano factor $1-C_2=F$ and $C_3$ as a function of system size.
The description of disorders applies to both figures. Dashed lines are 
theoretical diffusive values from Table 1.} 
\label{fig:1}
\end{figure}

$d=2$ is lower critical dimension for Anderson model
and the case with orthogonal symmetry has still no phase transition.
Any positive disorder forces that conductivity tends to zero for very
large systems. The obvious inequalities $0 < T_{M+1} < T_M < ... < T_1=G$
do not mean, that $C_M$ should necessarily tend to zero. Nevertheless this
can be expected for extremely large samples, as they approach the localized
(Poissonian) regime. From Fig. \ref{fig:1} we see, that $C_2, C_3$ are 
monotonic - decaying - functions of $L$ and that the theoretical metallic 
values are close to the plateau-like region, where the $L$-dependence is 
weak. But there is neither any special disorder, nor an asymptotic value, 
that would be reached. The $C_4, ... ,C_8(L)$ dependences look very similarly, 
just as if they were scaled down to lower values. The $C_M(1/L)$ dependences
for $W=3$ and $M=2,...,8$ in 2D were published already in \cite{pendry} 
(and compared to Eq. (\ref{eq:A2}) in \cite{bebu}), together with 3D data 
for $W=10$ and $W_C=16.5$.

\begin{figure}[t]
\noindent
\includegraphics[width=.4\textwidth]{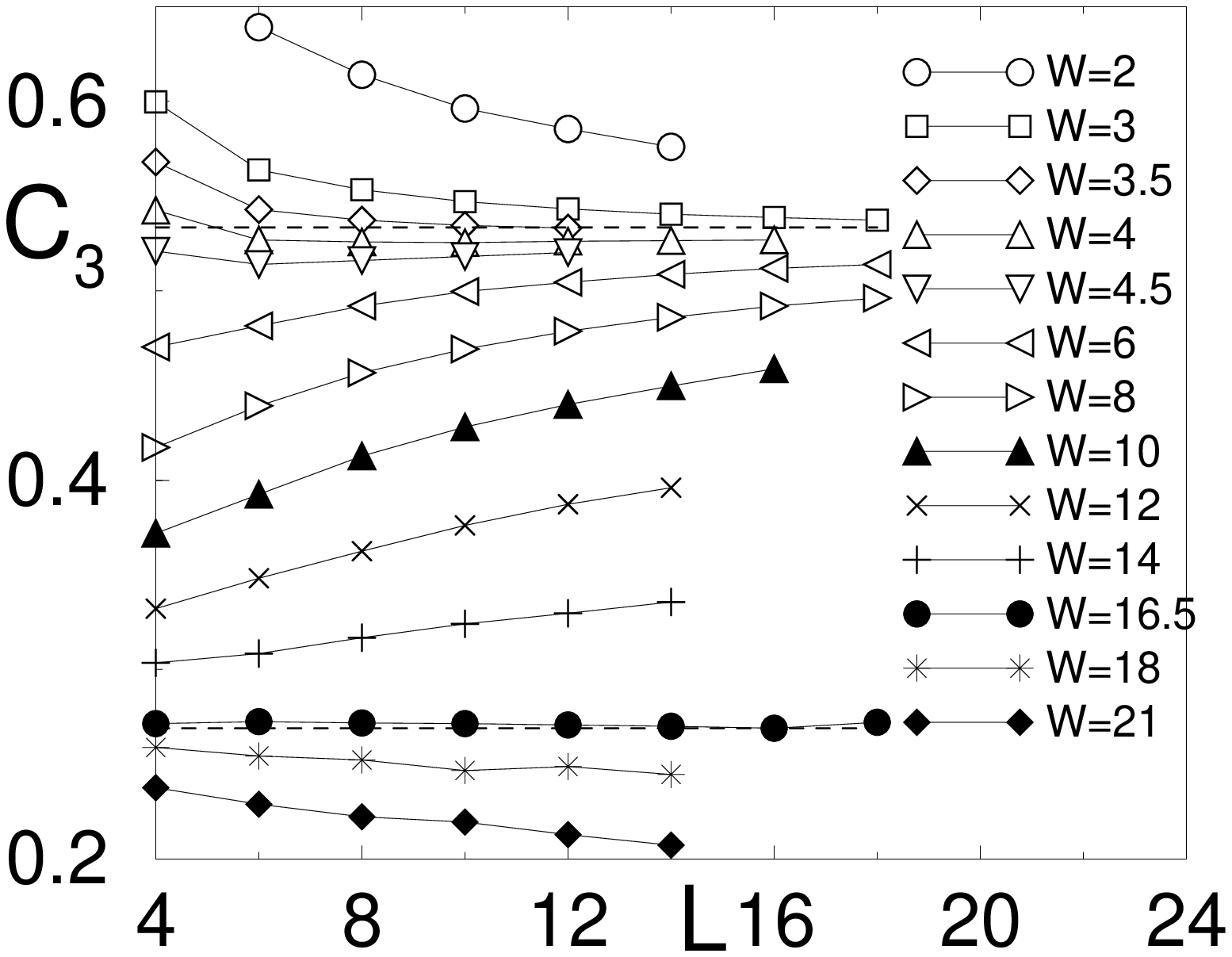}
\includegraphics[width=.395\textwidth]{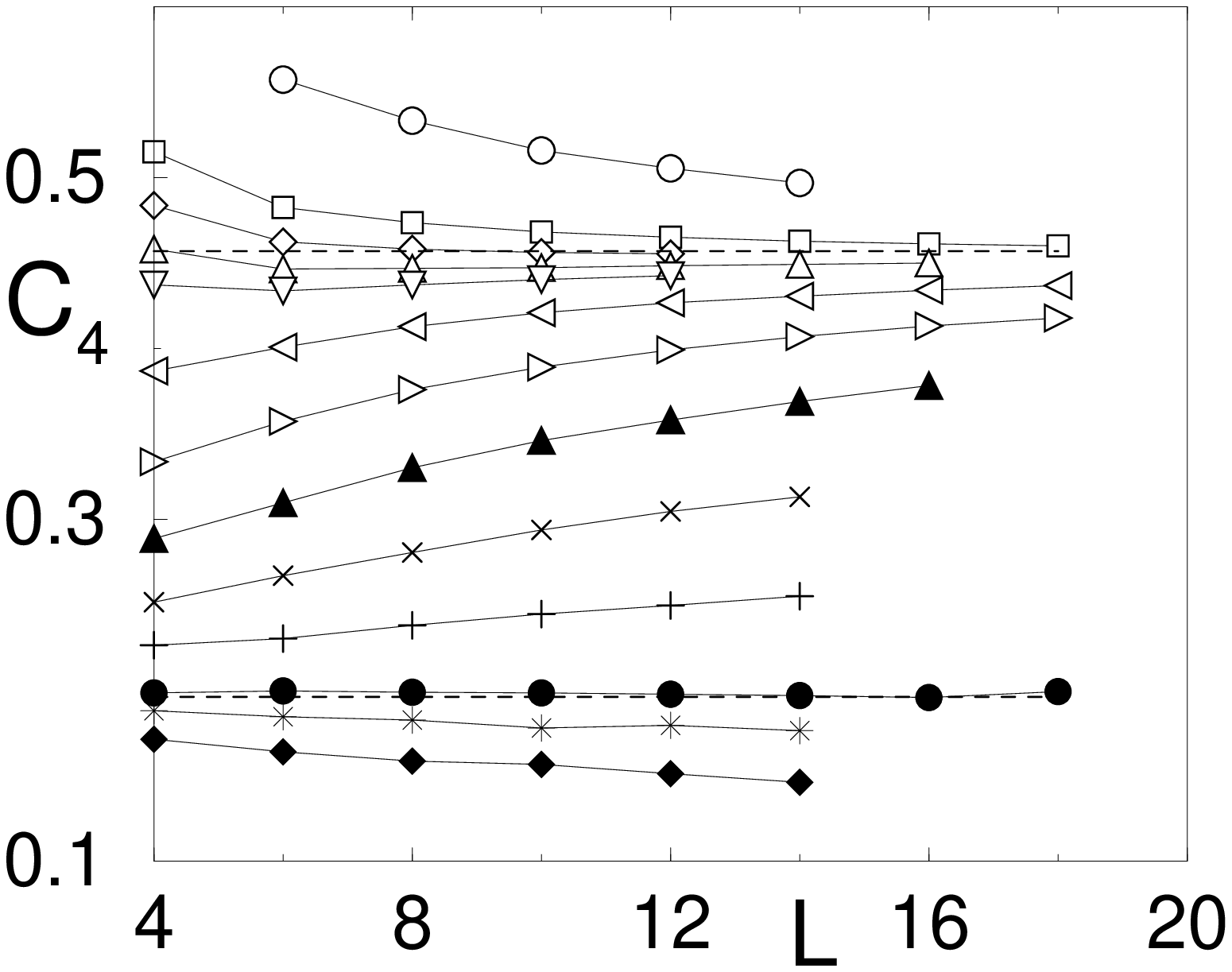}
\caption{$C_3$ and $C_4$ as a function of system size in 3D.
The description of disorders applies to both figures. Dashed lines are 
theoretical values from Table 1.}
\label{fig:2}
\end{figure}

\begin{figure}[t]
\noindent
\includegraphics[width=.4\textwidth]{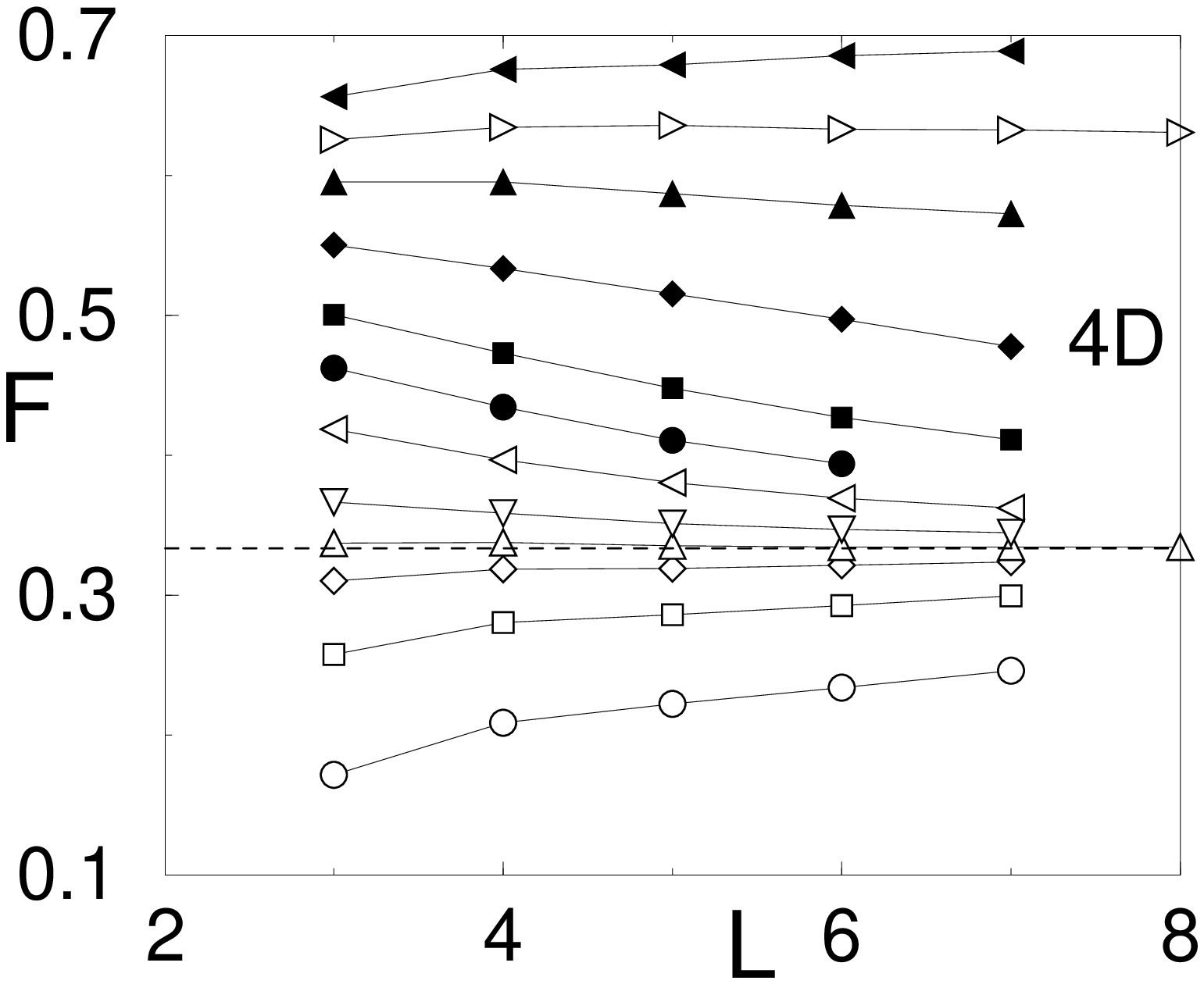}
\includegraphics[width=.39\textwidth]{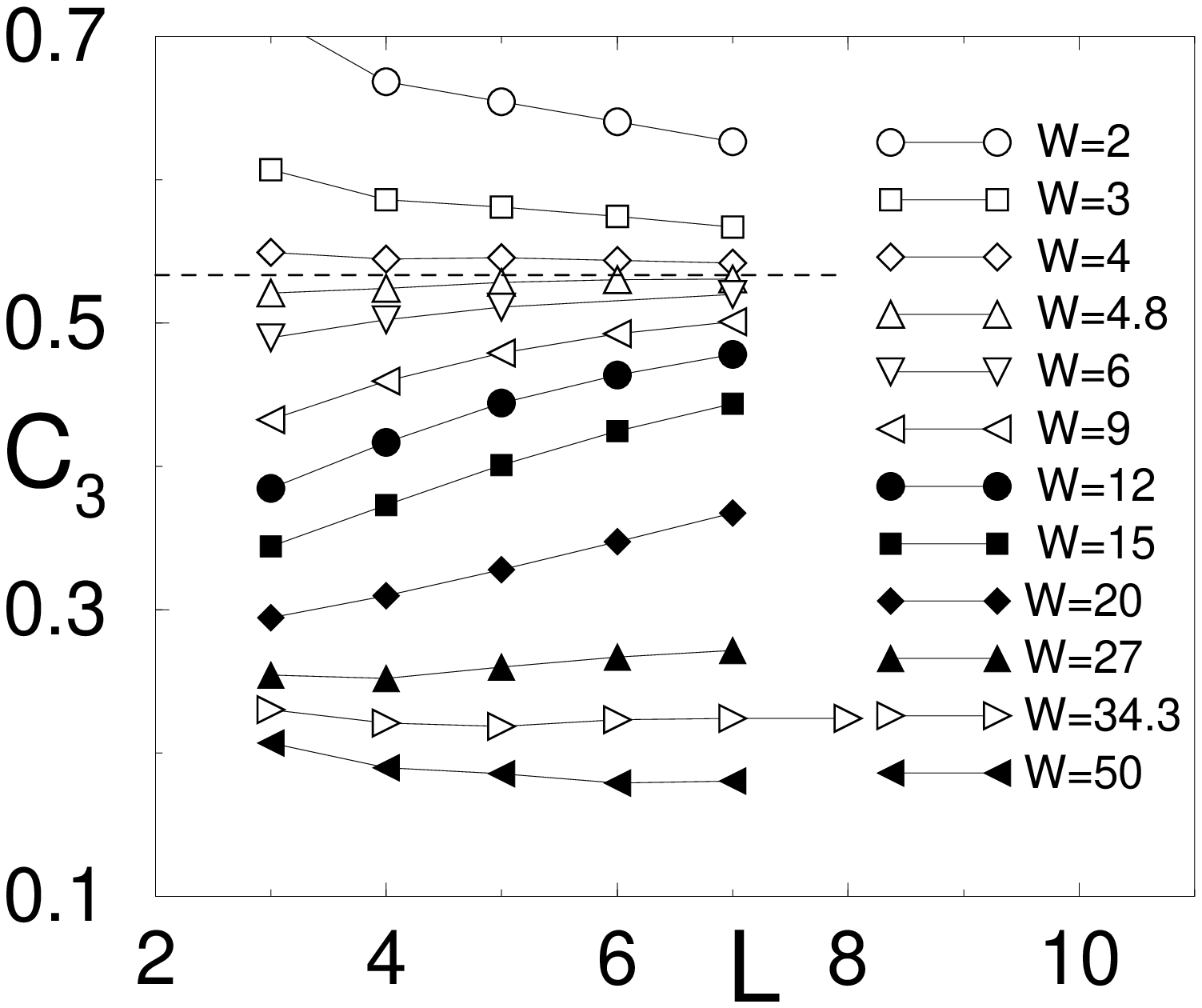}
\caption{Fano factor $F=1-C_2$ and $C_3$ as a function of system size in 4D.
Dashed lines are theoretical metallic values from Table 1. The description of 
disorders applies to both figures.}
\label{fig:3}
\end{figure}

The situation changes substantially in higher dimensions $d > 2$. 
The metallic limit becomes attractive for disorders $0 < W < W_C$.
The MIT is manifested by a constant line at $W=W_C$, which repels neighboring
lines on both sides, i. e. also for $W > W_C$, slowly approaching Poissonian
limit. This is why we think that any $C_M$ is a good order parameter,
the instability of the critical point follows from the scaling hypothesis.
In 3D both the metallic and critical values agree very well
with theoretical predictions of Eq. (\ref{eq:7}) for $\alpha = 0$ and 1, 
respectively, see Fig. \ref{fig:2}. Fano factor $F(L)$ has already been 
published in \cite{shot3}.  Concerning 4D, metallic limits are still the same,
as expected, see Fig. \ref{fig:3}. But the MIT values are not so easy 
to understand. We tried several conjectured $P(\lambda)$, e. g. with 
non-integer $\alpha$, or a polynomial. None of them could satisfy all $C_M$ 
values with low number of fitting parameters, though polynomials of the type
with $P_C(\lambda)=b_1\lambda+b_2\lambda^2+b_3\lambda^3$ can fit the moments 
within several percent. Most probably we have to do with some nontrivial 
function, too complicated to be found numerically. Not to speak about
the possibility of different $P_C(\lambda_1)$.
Of course, in 4D we can still have huge finite size effects, spoiling 
the precision of $C_M$ values at MIT. We leave the 4D $P_C(\lambda)$ 
question open for future investigations.

\subsection{Diffusive $R_M$ combinations of $T_M$}

\begin{figure}[t]
\noindent
\includegraphics[width=.4\textwidth]{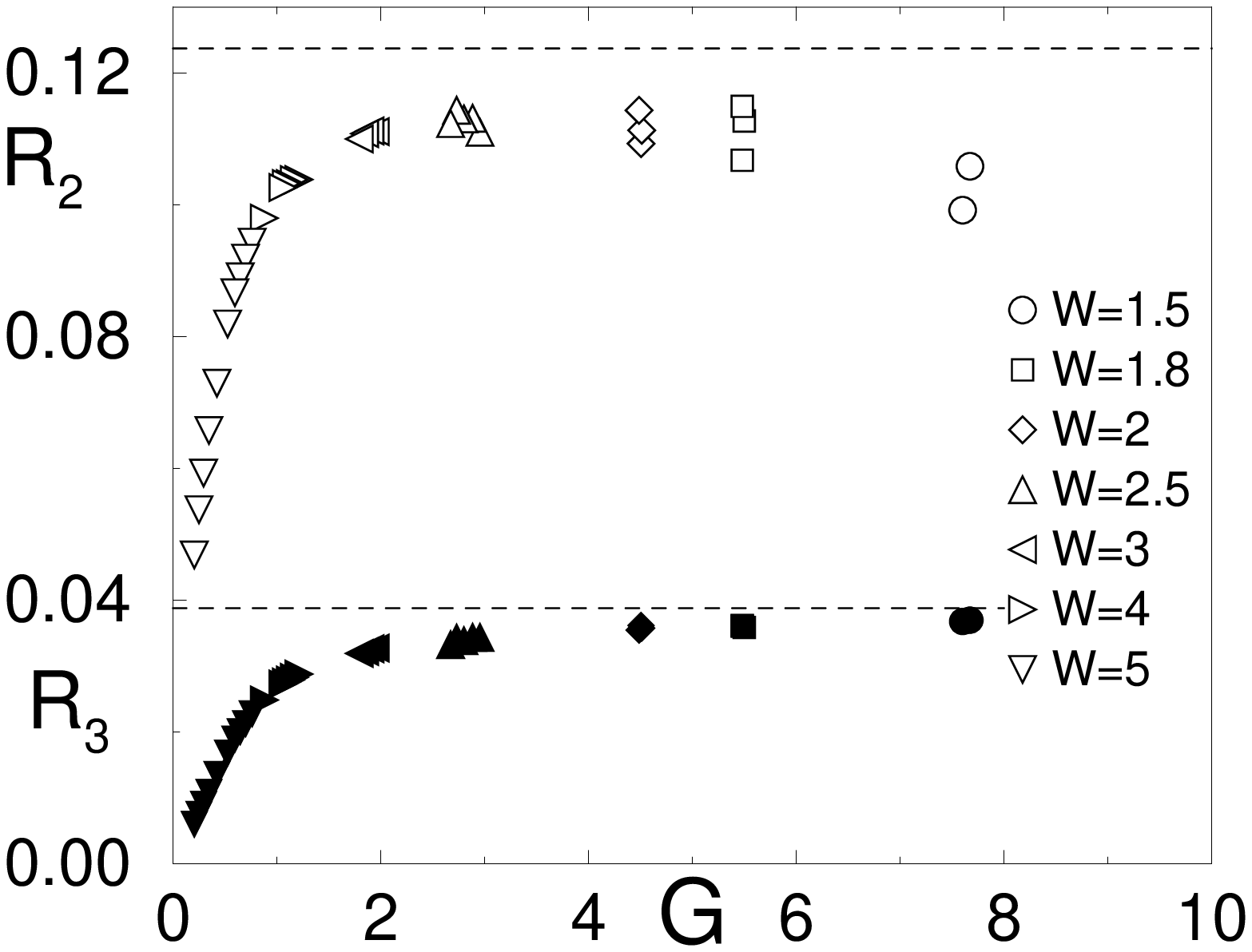}
\includegraphics[width=.395\textwidth]{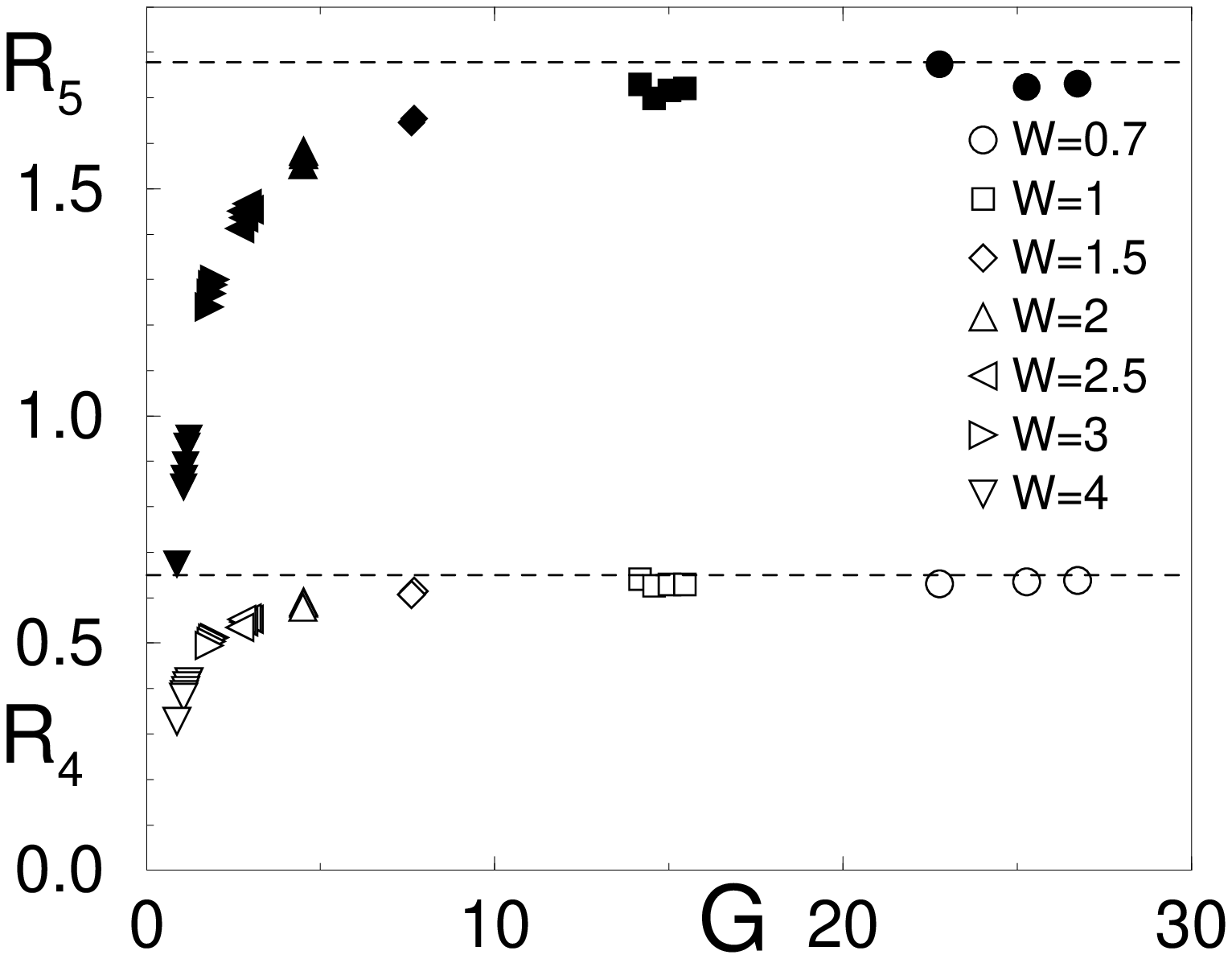}
\caption{Various $R_M$ quantities as a function of conductivity 
in 2D. Open symbols are $R_2$ and $R_4$, full ones $R_3$ and $R_5$.
Dashed lines are theoretical values from Table 2.}
\label{fig:4}
\end{figure}

\begin{figure}[t]
\noindent
\includegraphics[width=.4\textwidth]{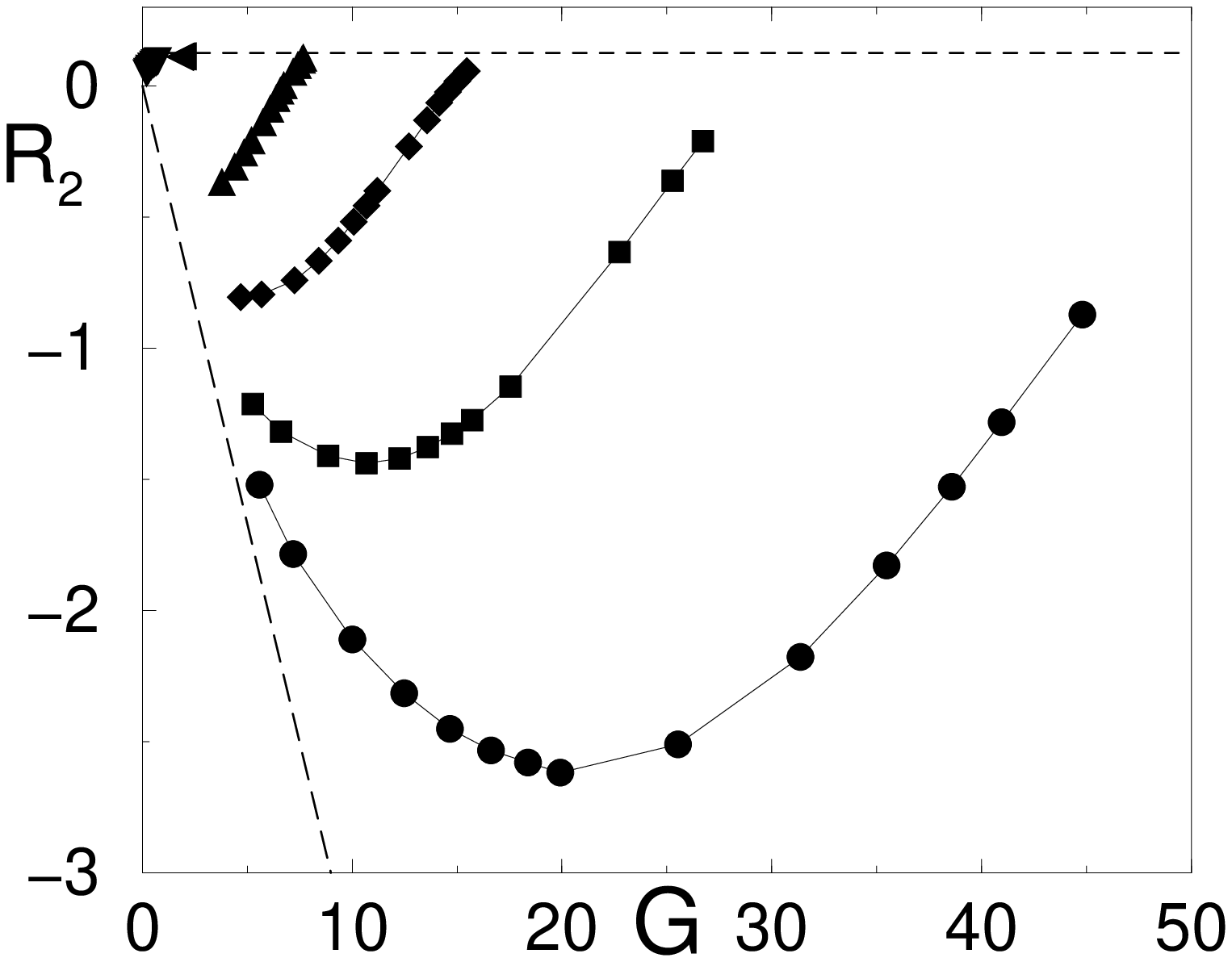}
\includegraphics[width=.41\textwidth]{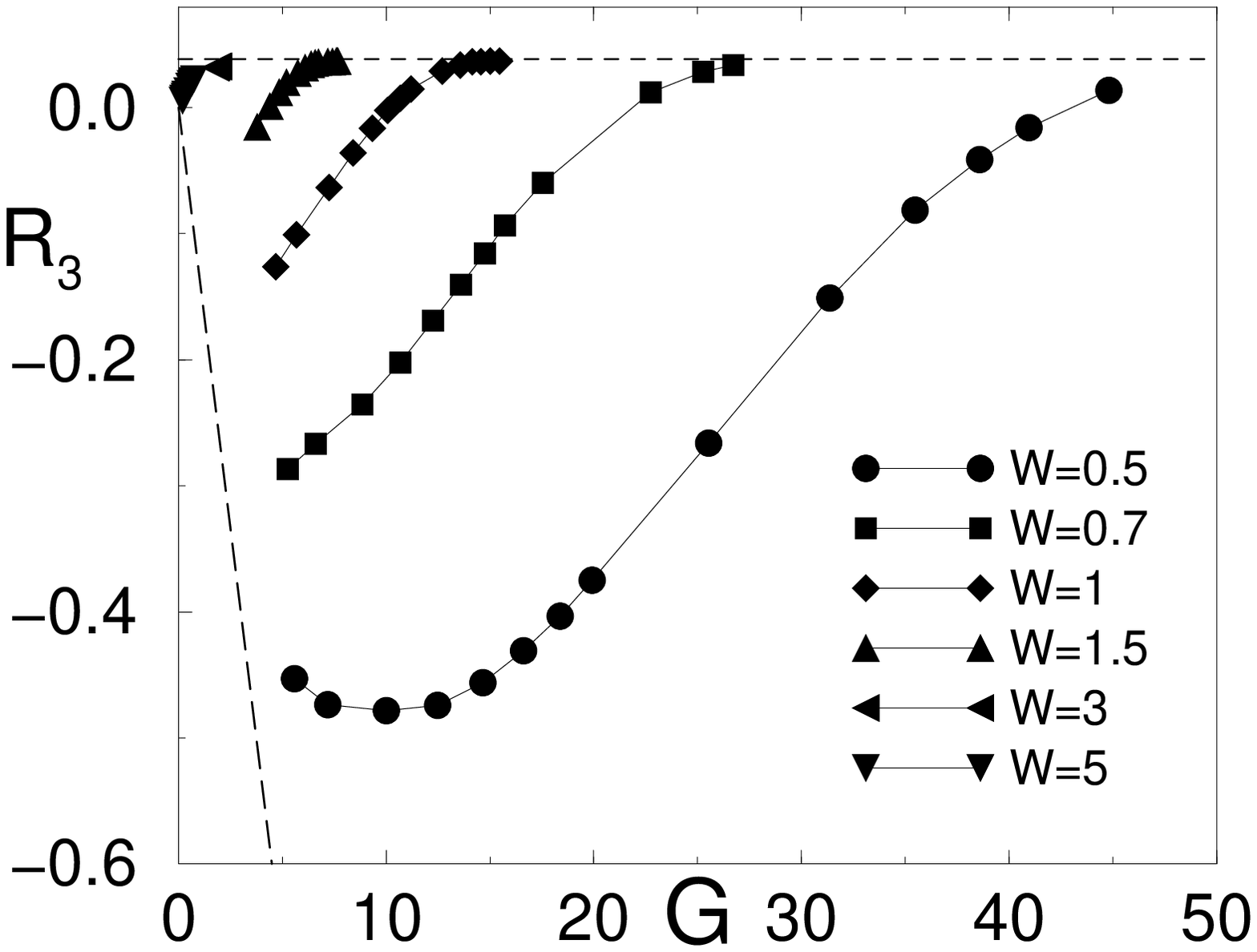}
\caption{ $R_2$ and $R_3$ in 2D including smaller $W$ and $L$.
Horizontal dashed lines are metallic theoretical values from Table 2. 
Long - dashed lines are ballistic limits. The description of disorders 
applies to both figures.}
\label{fig:5}
\end{figure}

We calculated also the averaged $R_M$ quantities in order to check Eq. (\ref
{eq:9}) numerically. It is usefull to plot them as a function of conductivity
$G$, rather than of $L$. Fig. \ref{fig:4} left shows, that in 2D the $R_2$ 
only roughly approaches the predicted value, at larger $G$ (smaller $W$) the
ballistic regime disturbs the convergence. $R_2(G)$ was shown already 
in \cite{shot12}. In order to judge the influence of ballistic, we plot
the data also for lower $W$ and $L$ in Fig. \ref{fig:5}. Without any disorder
($W=0$) all ${\cal T}_n=1$ and thus all $T_M=T_1$. So we get the ballistic
asymptotics, e. g. $-G/3$ and $-2G/15$ for $R_2$ and $R_3$, resp. One can 
see that the data for low $W$ and $L$ are close to the ballistic asymptotics
and with rising $L$ they approach the metallic limit. Surprisingly, for 
the higher indexed $R_3, R_4, R_5, ...$ this problem becomes less sensitive. 
Especially if we limit ourselves only to data for the several largest $L$, 
their behaviour is almost perfect, Fig. \ref{fig:4}, within few percent 
from theoretical values, Table \ref{tab:2}. To overcome the ballistic 
influence for $R_2$ fully, we would need even greater $L$ than available.

\begin{figure}[t]
\noindent
\includegraphics[width=.4\textwidth]{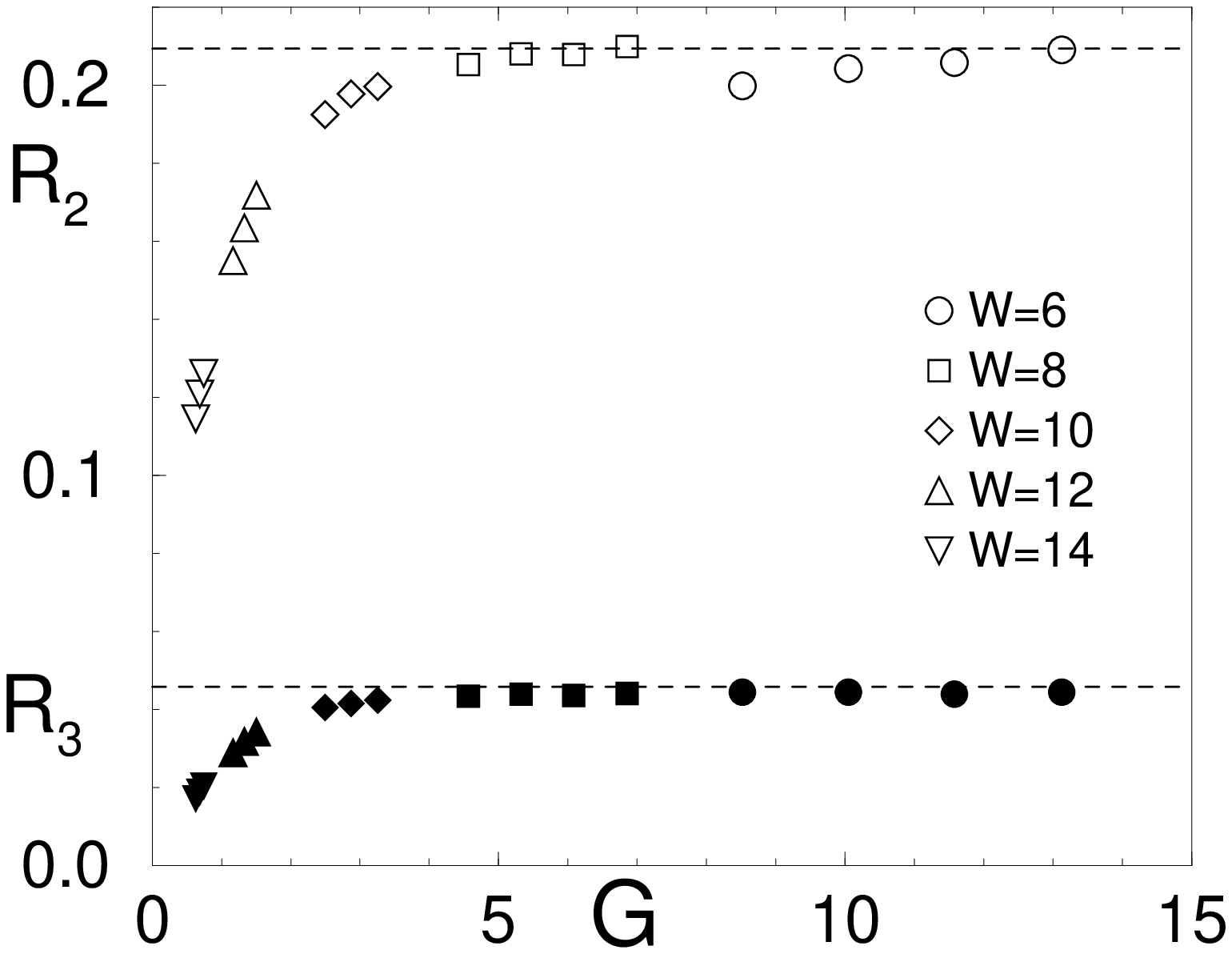}
\includegraphics[width=.395\textwidth]{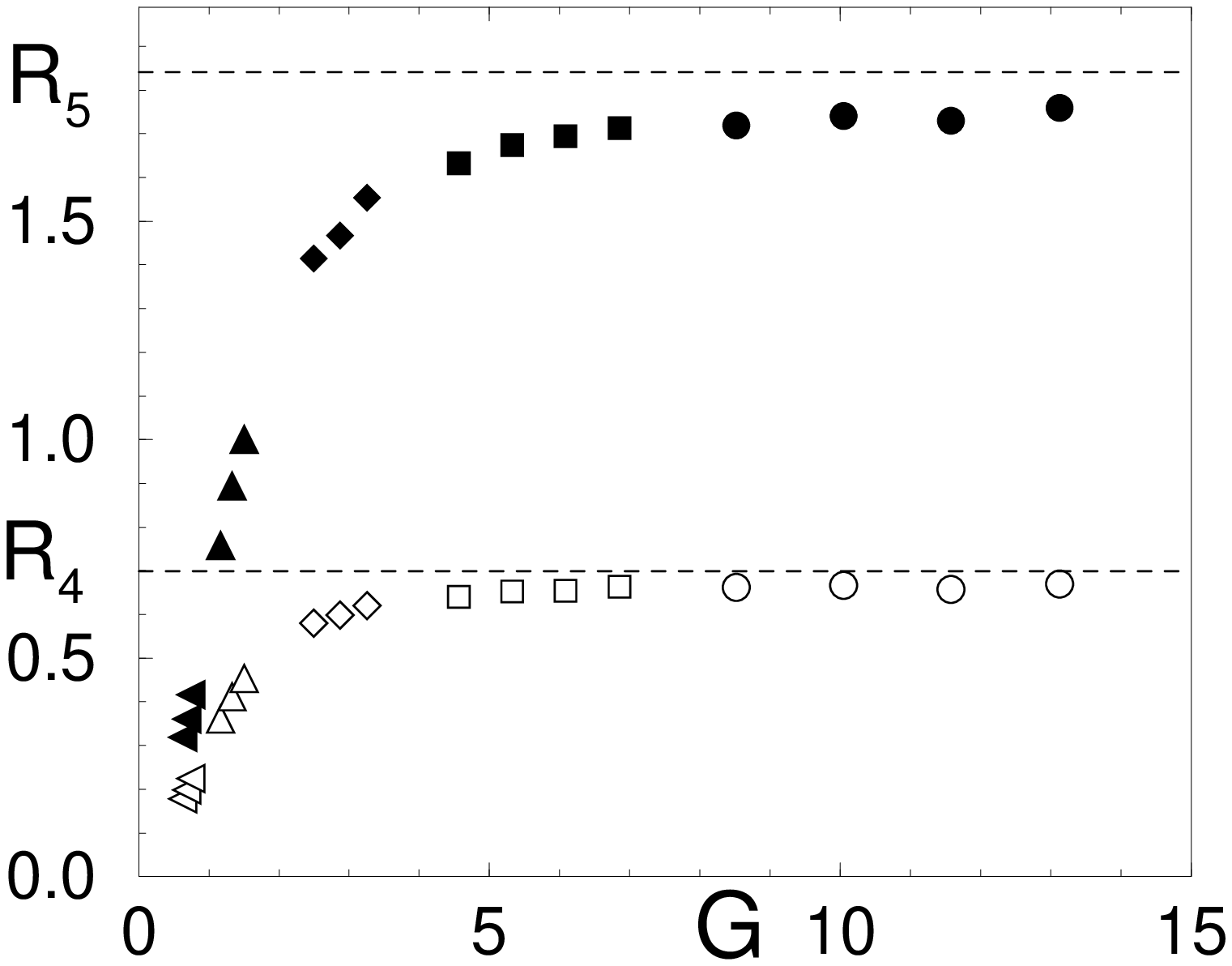}
\caption{Various $R_M$ quantities as a function of conductivity 
in 3D. Open symbols are $R_2$ and $R_4$, full ones $R_3$ and $R_5$.
Dashed lines are theoretical values from Table 2. The description 
of disorders applies to both figures.}
\label{fig:6}
\end{figure}

In 3D there are no signs of disturbing ballistics even for $R_2$ and our data
are in better accordance with predicted value than those in \cite{shot3}. 
Again the Eq. (\ref{eq:16}) is valid without any doubt, see Fig. \ref{fig:6}.

\begin{figure}[t]
\noindent
\includegraphics[width=.4\textwidth]{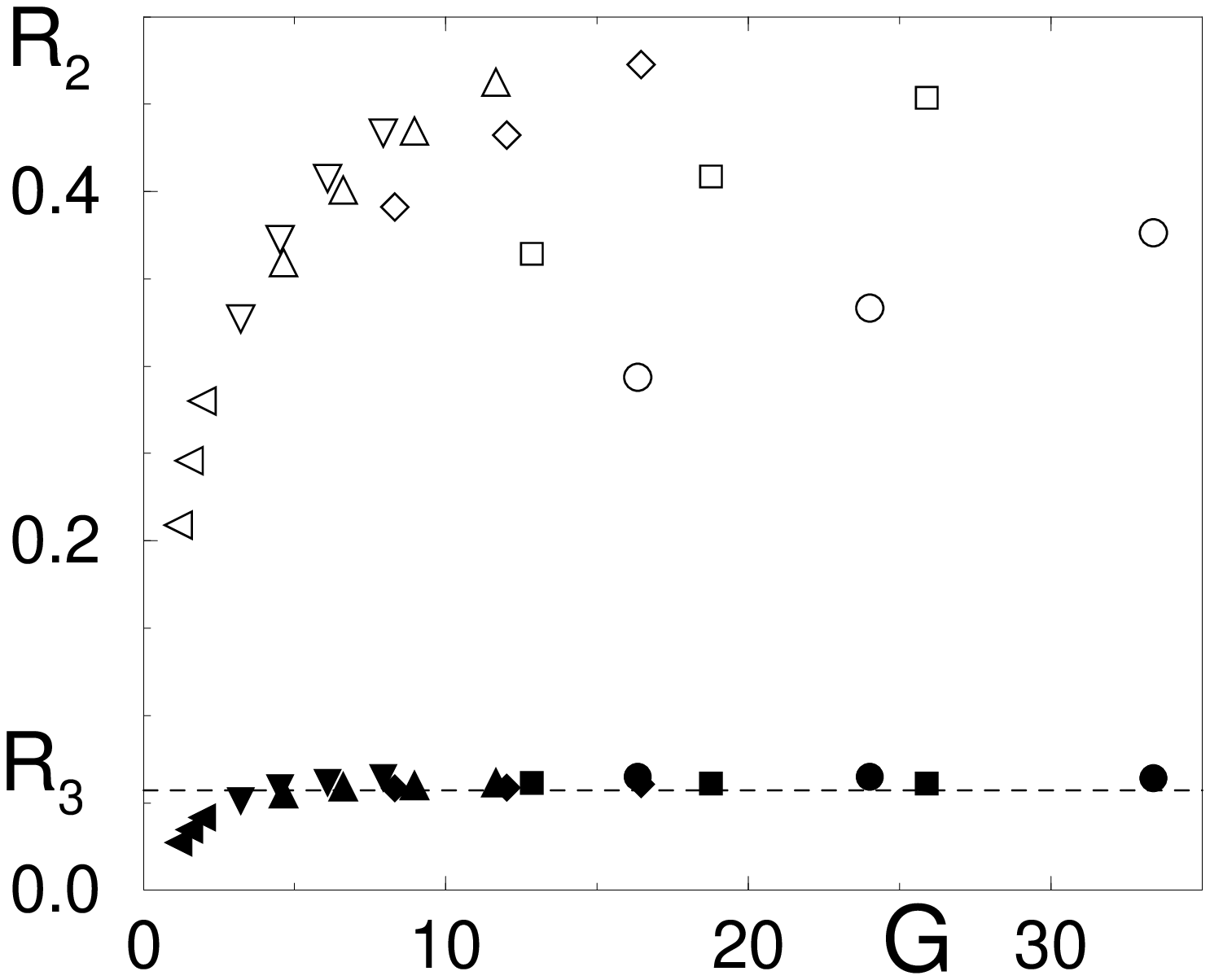}
\includegraphics[width=.4\textwidth]{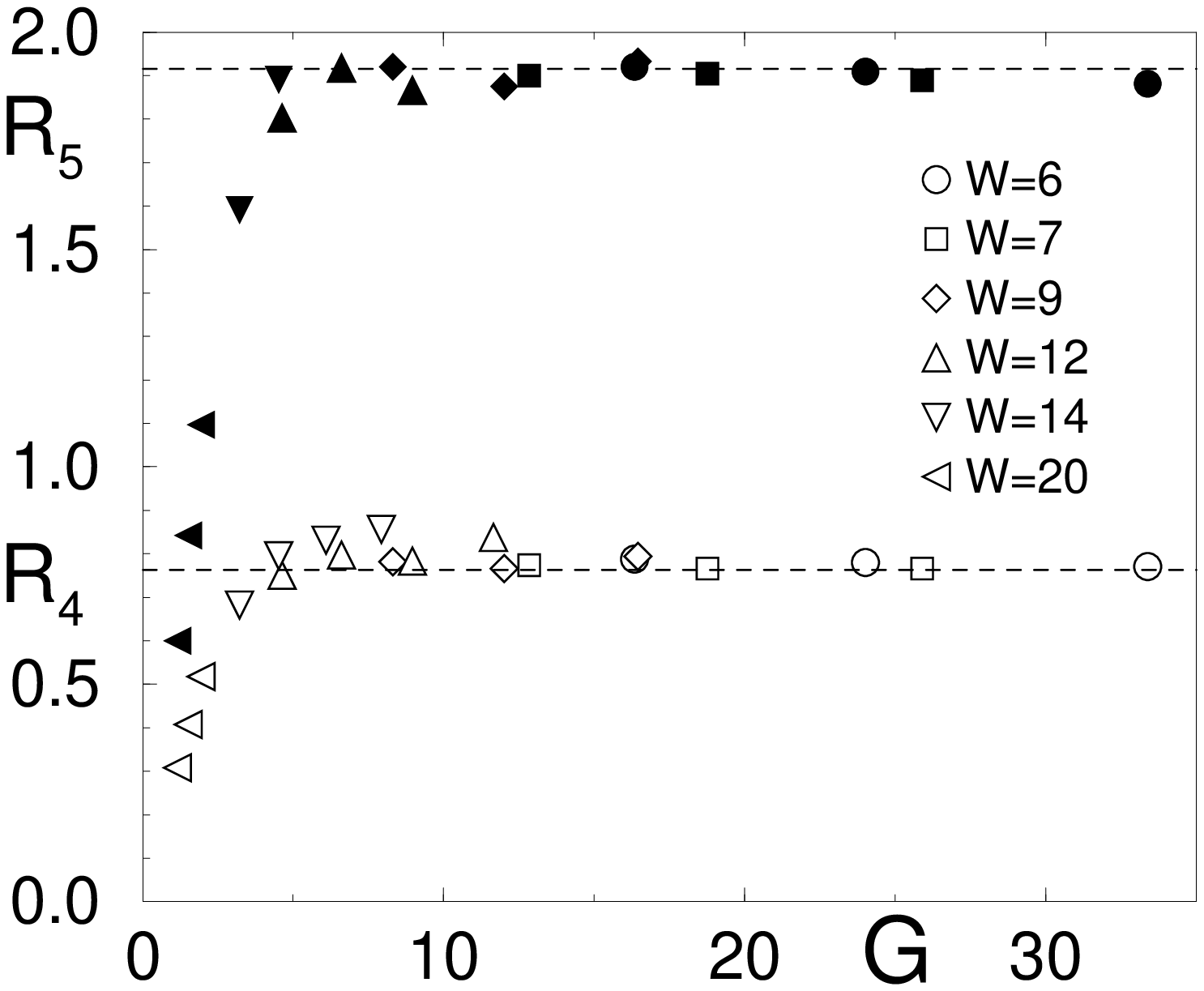}
\caption{Various $R_M$ quantities as a function of conductivity 
in 4D. Open symbols are $R_2$ and $R_4$, full ones $R_3$ and $R_5$.
Dashed lines are theoretical values from Table 2. The description 
of disorders applies to both figures.}
\label{fig:7}
\end{figure}

The $R_2$ in 4D is expected to diverge. This is not easy to show numerically,
as available system sizes are rather small, Fig. \ref{fig:7} left, and
ballistic influence is large again, similarly to 2D. Retrospectively
we might even say, that in 2D the $R_2$ alone was insufficient to confirm any
statement. Back in 4D we can just state, that simulated data are in no 
contradiction with possible logarithmic divergence $R_2 \propto \ln(G) 
\propto \ln(L)$, which we would expect from analogy with 4D UCF, see 
\cite{it}. On contrary, $R_3, R_4 ...$  converge again unexpectedly well to 
the predicted limits, Fig. \ref{fig:7}.

Summarizing, we regard Nazarov's formula (\ref{eq:9}) as confirmed 
by numerical simulations in Anderson model at least for dimensions up to 4.

\section{Conclusions}

We have shown all normalized higher current moments $C_M$ can be 
calculated exactly in the large system size limit at least in next regimes: 
the diffusive one in virtually any dimension \cite{bebu}, and the critical 
one (MIT) in 3D. We have confirmed the predicted values by numerical 
simulations. The MIT in 4D remains an open question. Further we saw, that 
Nazarov's formula (\ref{eq:16}) yields the same diffusive $C_M$ values as 
our Eq. (\ref{eq:A2}) and also special linear combinations of the $T_M$ 
moments converging to nontrivial constants. We have convincingly confirmed 
the validity of Eq. (\ref{eq:16}) for Anderson model in metallic regime by 
numerical simulations.

\section{Appendix A}

The aim is to find analytical solution of the integral

\begin{equation}
 \label{eq:A1}
 I_M(\alpha)=\int_0^\infty {x^\alpha \over \cosh^{2M}(x)} dx
\end{equation}

appearing in the Eq. (\ref{eq:5}) for $C_M=I_M/I_1$.
Special cases known from literature \cite{grry} are $\alpha=0$ (yielding
directly the $C_M(0)$ as $I_1(0)=1$, see \cite{bebu}):

\begin{equation}
 \label{eq:A2}
I_M(0)=C_M(0)=4^{M-1}{\rm B}(M,M)=\frac{1}{2}{\rm B}(1/2, M)=\frac{\Gamma(1/2)
\Gamma(M)}{2\Gamma(M+1/2)}
\end{equation}

and M=1:

\begin{equation}
 \label{eq:A3}
I_1(\alpha)=2^{1-\alpha}(1-2^{1-\alpha})\Gamma(\alpha+1)\zeta(\alpha)
\ \ \ \ {\rm if}\ \alpha\ne 1
\end{equation}

We introduced the special functions $\Gamma$, B and $\zeta$, i. e.
the Gamma, Beta and Rieman's Zeta function, respectively, \cite{grry}. 
Further we use
 the polylogarithmic function ${\rm Li}_{\alpha}(x)$:

\begin{equation}
 \label{eq:A4}
 {\rm Li}_{\alpha}(x)=\sum_{k=1}^{\infty}{x^k\over k^\alpha}
\end{equation}

more specifically its value at $x=-1$:

\begin{eqnarray}
 \label{eq:A5}
{\rm Li}_{\alpha}(-1)= & -(1-2^{1-\alpha})\zeta(\alpha) & {\rm if}\ \alpha \ne 1
\\                     & -\ln 2                         & {\rm if}\ \alpha = 1
\end{eqnarray}

Without going into details, we found the following exact formulas:

\begin{eqnarray}
 \label{eq:A6}
 I_1(\alpha) & = & -2^{1-\alpha}\Gamma(\alpha+1){\rm Li}_{\alpha}(-1) \\
 I_2(\alpha) & = & {2^{3-\alpha}\over 3!}\Gamma(\alpha+1)
[{\rm Li}_{\alpha-2}(-1)-{\rm Li}_{\alpha}(-1)] \\
 I_3(\alpha) & = & -{2^{5-\alpha}\over 5!}\Gamma(\alpha+1)
[{\rm Li}_{\alpha-4}(-1)-5{\rm Li}_{\alpha-2}(-1)+4{\rm Li}_{\alpha}(-1)] \\
 I_4(\alpha) & = & {2^{7-\alpha}\over 7!}\Gamma(\alpha+1)
[{\rm Li}_{\alpha-6}(-1)-14{\rm Li}_{\alpha-4}(-1)+49{\rm Li}_{\alpha-2}(-1)
\nonumber \\ & & -36{\rm Li}_{\alpha}(-1)] \\
 I_5(\alpha) & = & -{2^{9-\alpha}\over 9!}\Gamma(\alpha+1)
[{\rm Li}_{\alpha-8}(-1)-30{\rm Li}_{\alpha-6}(-1)+273{\rm Li}_{\alpha-4}(-1)
\nonumber \\ & & -820{\rm Li}_{\alpha-2}(-1)  +576{\rm Li}_{\alpha}(-1)] \\
 I_6(\alpha) & = & {2^{11-\alpha}\over 11!}\Gamma(\alpha+1)
[{\rm Li}_{\alpha-10}(-1)-55{\rm Li}_{\alpha-8}(-1)+1023{\rm Li}_{\alpha-6}(-1)
\nonumber \\ & &  -7645{\rm Li}_{\alpha-4}(-1) +21076{\rm Li}_{\alpha-2}(-1)
-14400{\rm Li}_{\alpha}(-1)]
\end{eqnarray}
It is worth mention that ${\rm Li}_{\alpha}(-1)$ and $\zeta(\alpha)$
have rational or zero value for $\alpha$ negative integer or zero.
In particular, $I_1(1)=\ln 2$, $I_2(1)=-\frac{1}{6}+\frac{2}{3}\ln 2$, 
$I_3(1)=-\frac{11}{60}+\frac{8}{15}\ln 2
$, $I_4(1)=-\frac{19}{105}+
\frac{16}{35}\ln 2$, etc. These rationals (the term at $\ln 2$ is $I_M(0)$) 
can be found much easier with the help of the generator \cite{grry}:

\begin{equation}
 \label{eq:A7}
I_G(t)=\int_0^\infty \frac{x\ dx}{\cosh x+\cos 2t}=4\frac
{t \ln 2 - L(t)}{\sin 2t}
\end{equation}
One just takes consecutive odd derivatives at $t=\pi/4$.
It is interesting to compare this generator to Eq. (\ref{eq:18}).  
Integrating both sides of Eq. (\ref{eq:A7}) w. r. to $y=-\sin^2 t$ 
yields Eq. (\ref{eq:10}). 

\section{Appendix B}

In order to get relations for $T_M$, we will perform $n$-fold derivation
$(n = 0,1, ..., M-1)$ of $\delta F(\phi)$  with respect 
to $\phi^2$ at $\phi=0$, on both terms from Eq. (\ref{eq:16}):
\begin{eqnarray}
 \label{eq:A8}
\lim_{\phi \to 0} \delta F & = & T_1 = -2S_1 \\
\lim_{\phi \to 0} {d \over d \phi^2}\delta F=\lim_{\phi \to 0}
{1\over 2 \phi} {d\over d\phi}\delta F & = & {T_2\over 4} =
-2S_2-{S_1\over 3}\\
\lim_{\phi \to 0} {d^2 \over (d \phi^2)^2}\delta F & = & {1\over 24}
(3T_3-T_2) = -4S_3-{2\over 3}S_2-{7\over 90}S_1
\end{eqnarray}
etc. We solve this linear system of equations so that we remove almost
all the sums $S_n$, except of the one with the highest index. Thus we get
the set of Eq. (\ref{eq:17}). We choose this one from the plenty of 
possibilities, because $S_M$ diverges for $d \ge 2M$, like $S_2$ and $R_2$ 
at 4D, Fig. \ref{fig:7}. But
$S_3$ is finite at 4D (even 5D) and it can be compared to numerical results.
The case $M=2$ was already treated this way in \cite{shot12} and 
\cite{shot3} and the sums $S_2(d)$ were calculated for $d=2,3$.

Now let us show how to calculate the $d$-fold sums $S_M(d)$ easily with
much better precision than by direct summation. The statement is:
\begin{equation}
 \label{eq:A9}
S_M(d)=\frac{1}{\pi^M 2^d}\int_0^{\infty}{y^{M-1}\over (M-1)!}[\vartheta_3(0,
e^{-\pi y})-1][\vartheta_3(0,e^{-\pi y})+1]^{d-1} dy
\end{equation}
Here $\vartheta_3$ is Jacobi elliptic function, simplified to:
\begin{equation}
 \label{eq:A10}
\vartheta_3(0,q)=\sum_{k=-\infty}^\infty q^{k^2}=\frac{1}{\sqrt y}
\sum_{k=-\infty}^\infty \exp\Bigl(-\frac{k^2 \pi}{y}\Bigr)
\end{equation}
with $q=\exp(-\pi y)$. The proof of Eq. (\ref{eq:A9}) consists in using 
the first expansion of elliptic functions and performing  per partes 
$(M-1)$-times. This time the case $M=2$ was proposed in \cite{it}, dealing 
with universal conductance fluctuations in arbitrary dimension.

Finally let us derive the convergence condition for $S_M(d)$. The values
of the expression under integral (\ref{eq:A9}) at small $y$ are crucial. We 
use the second expansion $\vartheta_3(0,q) \approx 1/\sqrt y$ (case $k=0$) 
and the leading term is proportional to $y^{M-1-d/2}$. The turning point 
of convergence is $y^{-1}$, i. e. $M=d/2$, with logarithmic type 
of divergence \cite{it}.

{\bf Acknowledgement}

We are grateful to P. Marko\v s for valuable comments, the VEGA 
grant Nr. 2/3108/23 for financial support and the Computer Centre 
of Slovak Acad. Sci. for CPU time.

\end{document}